\documentclass[iop]{emulateapj}
\usepackage{multirow}
\usepackage{bm}
\usepackage{amssymb}
\usepackage{amsmath}
\usepackage{latexsym}
\usepackage{graphicx}
\usepackage{ifsym}
\usepackage{bm}
\usepackage{tabularx}
\usepackage{hyperref}
\usepackage{enumitem}

\slugcomment{}
\shorttitle{Compact Optically-thick Disks}
\shortauthors{Wu et al.}

\begin{document}
\title{\textbf {\large A\lowercase{n} E\lowercase{xplanation of the} V\lowercase{ery} L\lowercase{ow} R\lowercase{adio} F\lowercase{lux of} Y\lowercase{oung} P\lowercase{lanet-mass} C\lowercase{ompanions}}}
\author{Ya-Lin Wu$^1$, Laird M. Close$^1$, Josh A. Eisner$^1$, and Patrick D. Sheehan$^2$}
\affil{$^1$Steward Observatory, University of Arizona, Tucson, AZ 85721, USA\\ \vspace{4pt}
$^2$Homer L. Dodge Department of Physics and Astronomy, University of Oklahoma, Norman, OK 73019, USA\\
{\it Accepted for Publication in AJ}}

\begin{abstract} 
We report Atacama Large Millimeter/submillimeter Array (ALMA) 1.3 mm continuum upper limits for 5 planetary-mass companions DH Tau B, CT Cha B, GSC 6214-210 B, 1RXS 1609 B, and GQ Lup B. Our survey, together with other ALMA studies, have yielded null results for disks around young planet-mass companions and placed stringent dust mass upper limits, typically less than 0.1 $M_{\oplus}$, when assuming dust continuum is optically thin. Such low-mass gas/dust content can lead to a disk lifetime estimate (from accretion rates) much shorter than the age of the system. To alleviate this timescale discrepancy, we suggest that disks around wide companions might be very compact and optically thick, in order to sustain a few Myr of accretion yet have very weak (sub)millimeter flux so as to still be elusive to ALMA. Our order-of-magnitude estimate shows that compact optically-thick disks might be smaller than 1000 $R_{\rm Jup}$ and only emit $\sim$$\mu$Jy of flux in the (sub)millimeter, but their average temperature can be higher than that of circumstellar disks. The high disk temperature could impede satellite formation, but it also suggests that mid- to far-infrared might be more favorable than radio wavelengths to characterize disk properties. Finally, the compact disk size might imply that dynamical encounters between the companion and the star, or any other scatterers in the system, play a role in the formation of planetary-mass companions. 
\end{abstract}


\section*{\textbf {\normalsize1. I\lowercase{ntroduction}}}
Direct imaging of circumplanetary disks is crucial to study how planets and their satellites actually form. Theoretical modeling has suggested that circumplanetary disks can emit significant infrared fluxes, and magnetospheric accretion onto the surface of protoplanets can create strong line emission and UV/optical continuum excess (e.g., \citealt{E15,Z15,Z16}). Recently, high-contrast imaging campaigns have been targeting transition disks to search for these accretion signatures inside the central holes or gaps. While a few red sources have been identified as protoplanet candidates, their nature remains elusive and some may actually be disk features or data reduction artifacts (e.g., \citealt{Biller14,S15a,F17}). So far the only confirmed accreting planet is LkCa 15 b, which emits at H$\alpha$ \citep{S15b}, but its accretion disk has yet to be directly resolved \citep{I14}. 

One can also search for disks around wide-orbit planetary-mass companions, which are $\sim$5--40 $M_{\rm Jup}$ gas giants orbiting at tens to hundreds of astronomical units. Wide companions, unlike the copious transiting planets, are intrinsically rare (e.g., \citealt{NC10}), so their formation scenario may not resemble that of planets (i.e., core accretion) but more like brown dwarfs or low-mass stars (e.g., \citealt{Brandt14}). Nonetheless, their wide separations offer a direct view into the physical mechanisms that can regulate planet formation in circumplanetary disks. Analytic and numerical analyses have shown that circumplanetary disks are truncated at $\sim$1/3 of the planet's Hill radius (e.g., \citealt{QT98,AB09}). \cite{SB13} further demonstrated that gas giants formed via disk fragmentation at 100 au can harbor such a truncated disk. Simulations have predicted that circumplanetary disks are luminous at radio wavelengths and may be easily detectable by the Atacama Large Millimeter/submillimeter Array (ALMA) (e.g., \citealt{SB13,Z16}). 

\begin{figure*}[t]
\centering
\includegraphics[angle=0,width=0.33\linewidth]{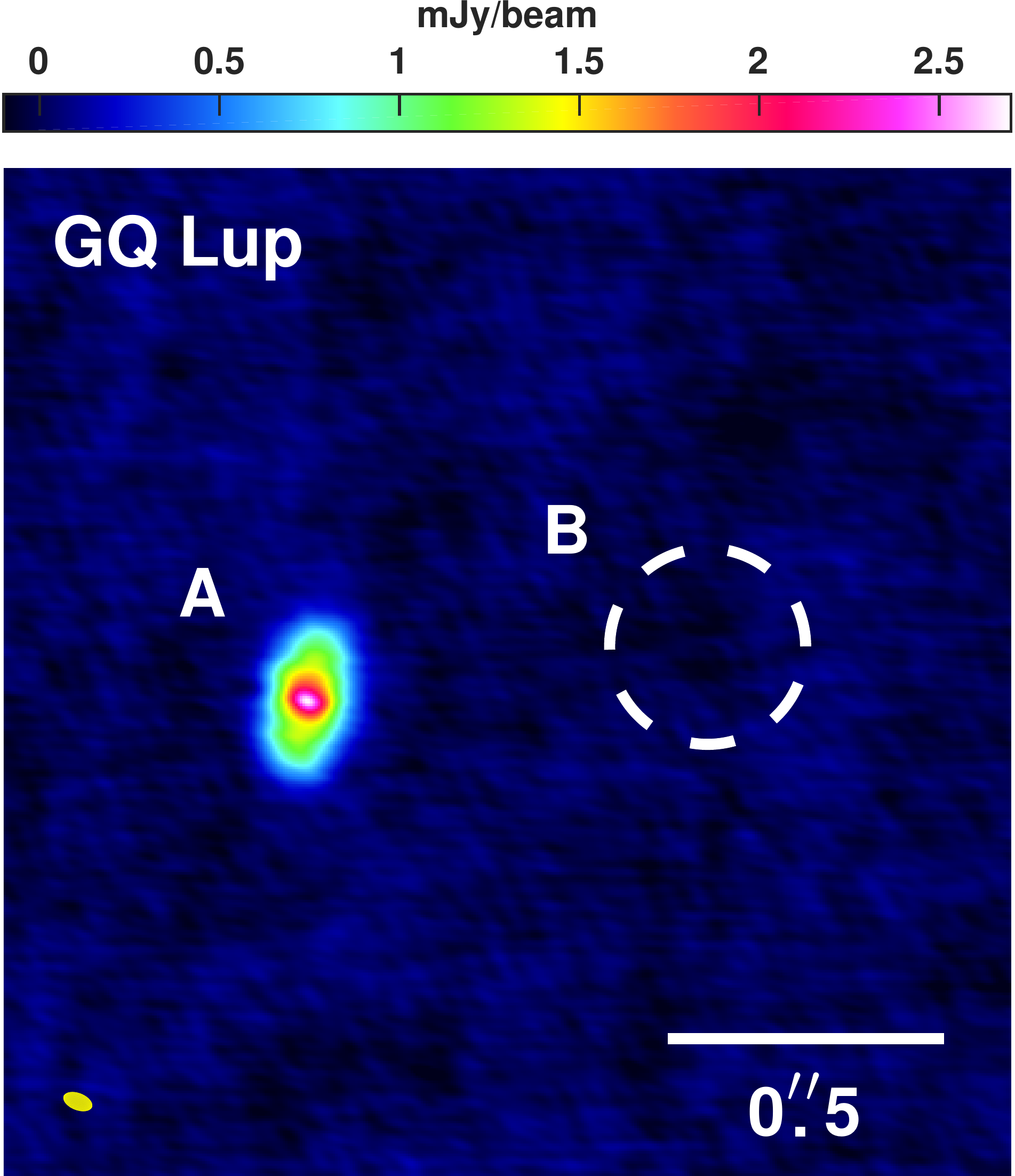}
\includegraphics[angle=0,width=0.33\linewidth]{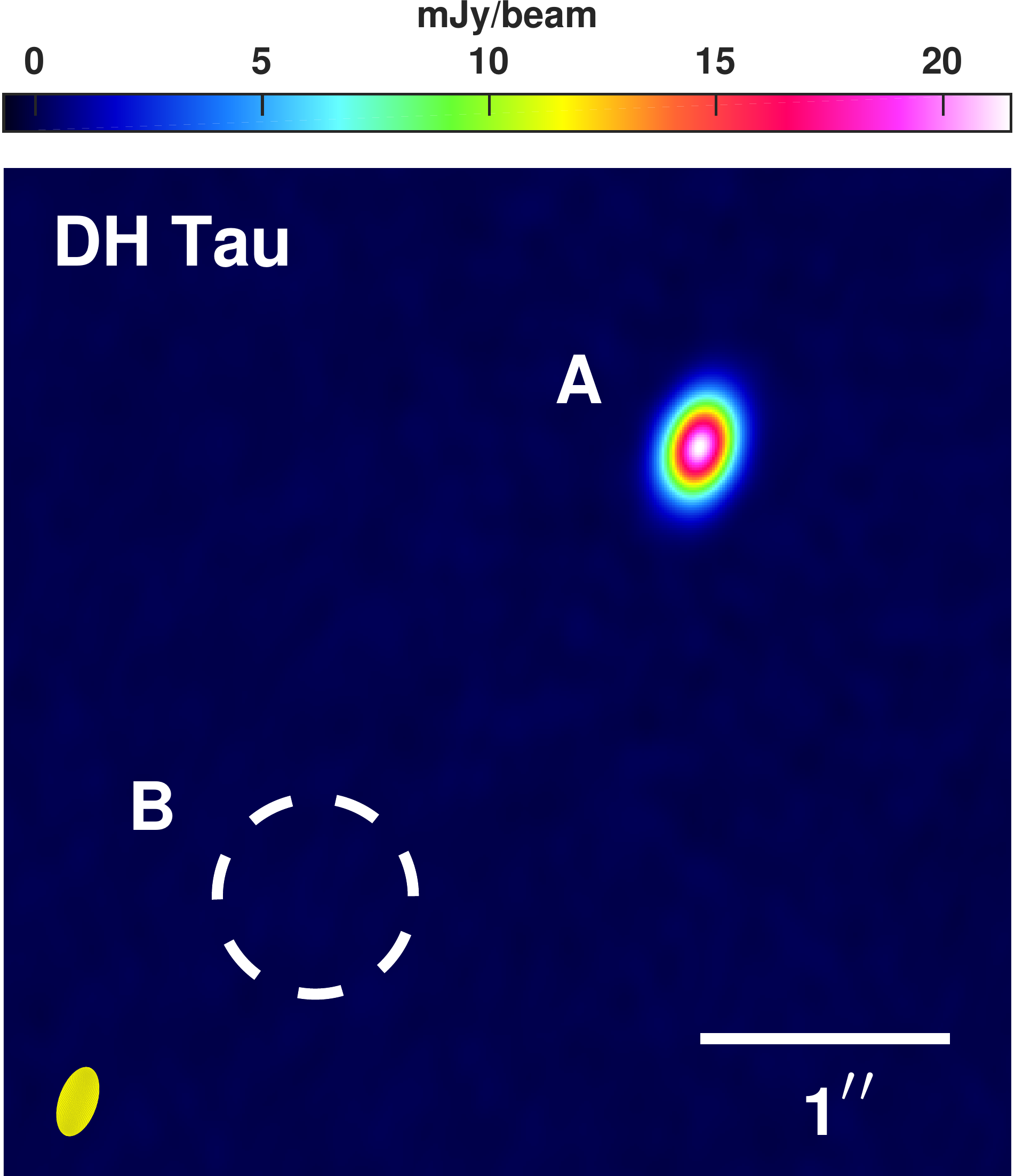}\vspace{6pt}\\
\includegraphics[angle=0,width=0.33\linewidth]{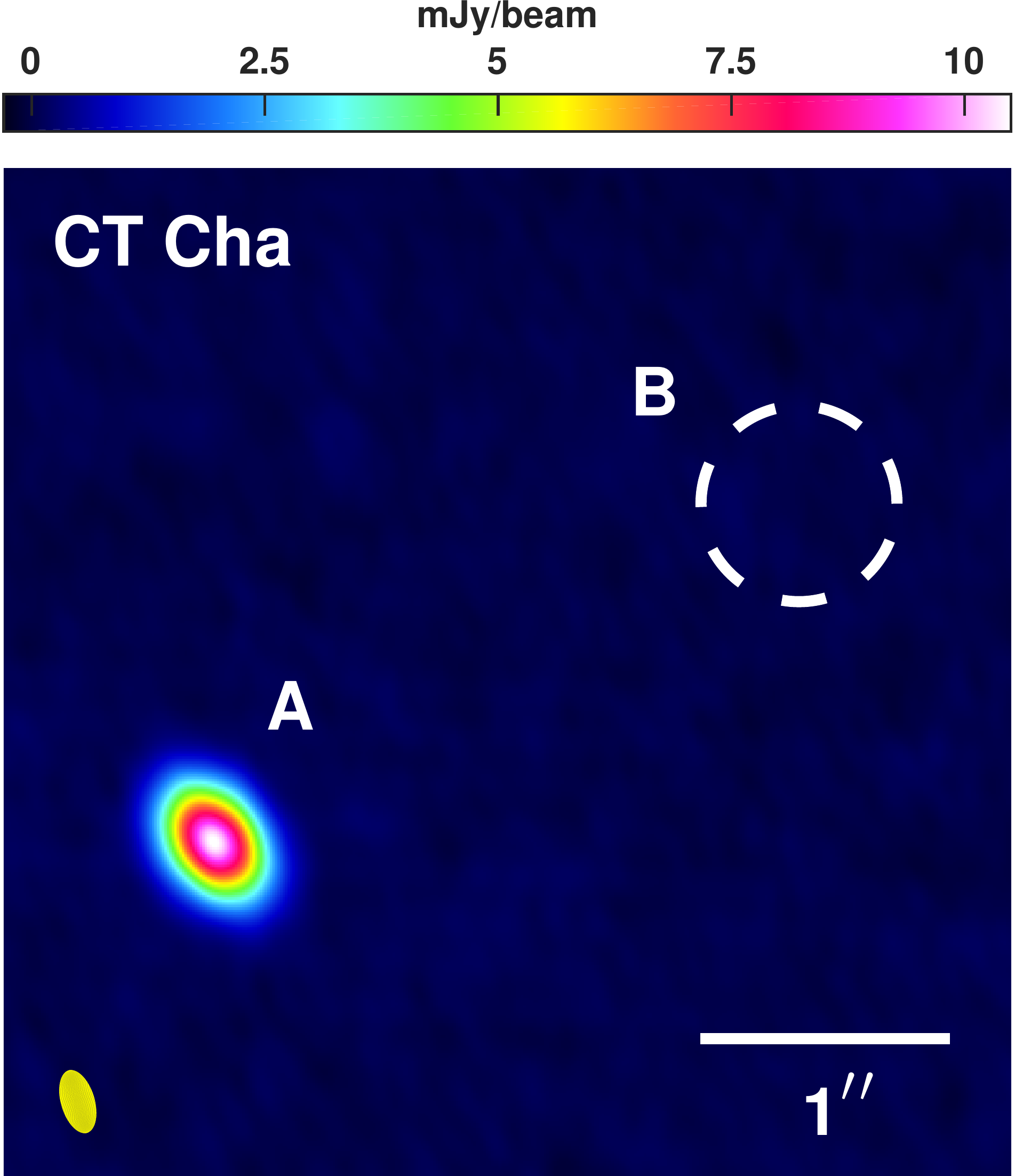}
\includegraphics[angle=0,width=0.33\linewidth]{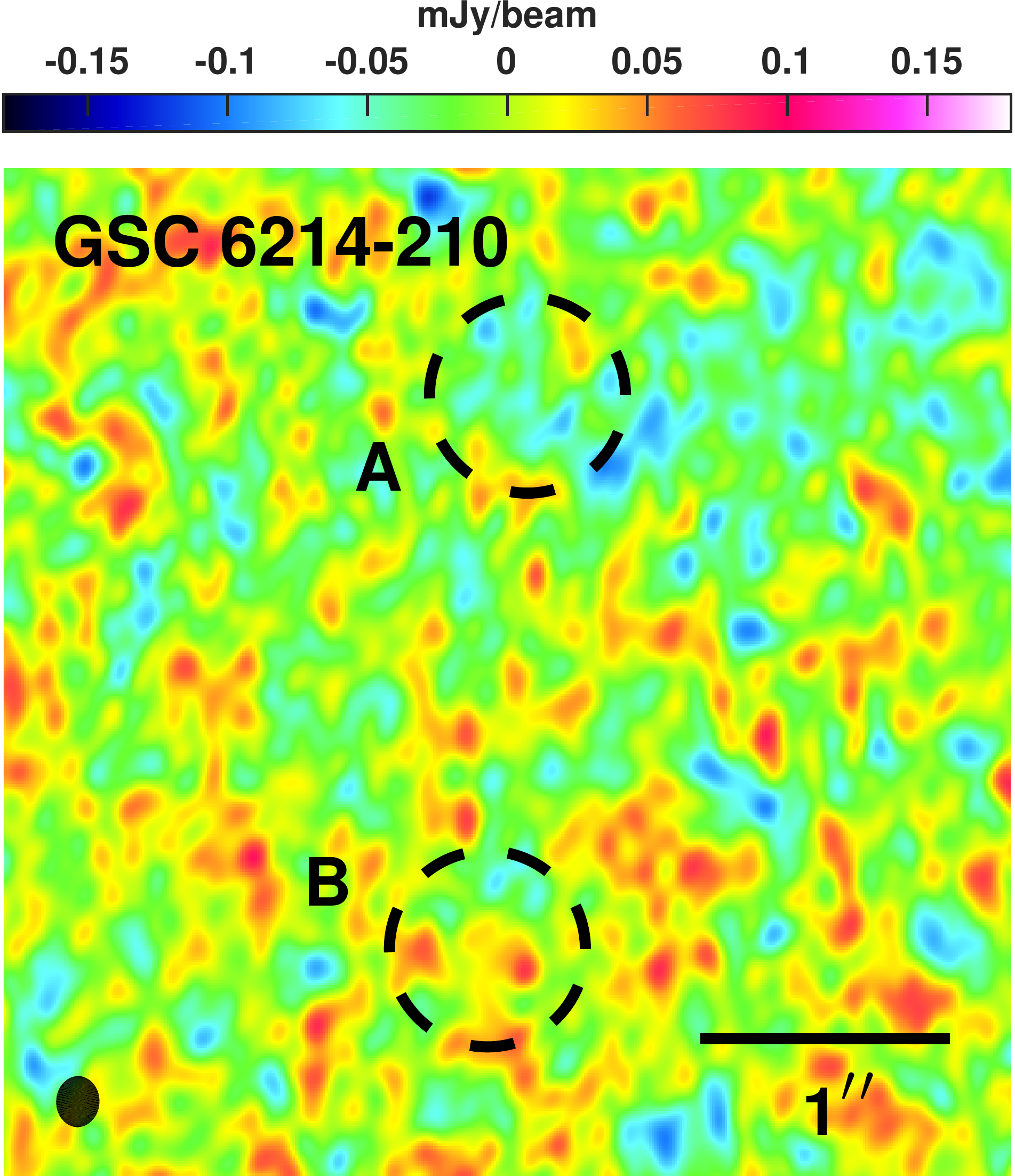}
\includegraphics[angle=0,width=0.33\linewidth]{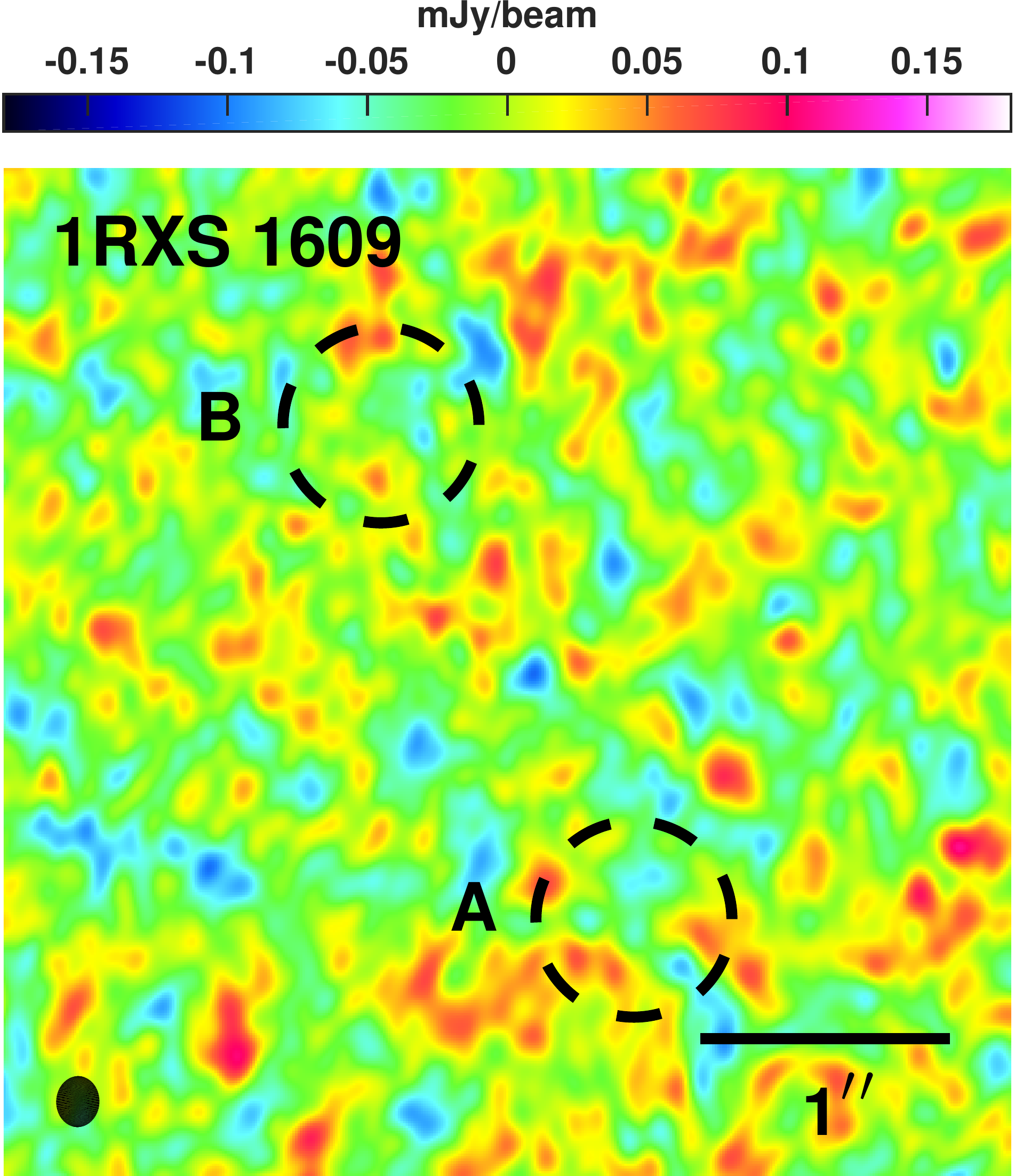}
\caption{A gallery of ALMA 1.3 mm non-detections. The 3$\sigma$ flux limits range from 90 to 150 $\mu$Jy, corresponding to 0.06 to 0.14 $M_\oplus$ of dust under optically thin approximation. North is up and east is left for all panels. We detect circumstellar disks around GQ Lup A, CT Cha A, and DH Tau A, with fluxes of $\sim$27, $\sim$35, and $\sim$33 mJy and disk radii of $\sim$20, $\sim$41, and $\sim$16 au, respectively. The GQ Lup A and CT Cha A disks are spatially resolved, while the DH Tau A disk is marginally resolved. Full analysis of the GQ Lup A disk is reported in \cite{W17}, and disk modeling for CT Cha A and DH Tau A will be presented in a future paper.}
\label{fig0}
\end{figure*}

Recently, there have been several attempts to search for accretion disks around wide companions with radio interferometers such as the NOrthern Extended Millimeter Array (NOEMA) and ALMA (\citealt{B15, M17, W17, Wolff17, R17}). None of these studies has successfully detected a wide companion disk. In Table \ref{tb0} and Figure \ref{fig0}, we show our ALMA 1.3 mm survey for 5 systems, GQ Lup\footnotemark[1], CT Cha, DH Tau, GSC 6214-210, and 1RXS 1609. Data have been self-calibrated and CLEANed with natural weighting. Details on the observational setup and data reduction can be found in \cite{W17} and \cite{WS17}. As with other studies, no companions are detected in our survey. We note that a disk has been detected around the proposed planet-mass companion FW Tau C \citep{K14,K15,C15}, but a recent dynamical mass measurement has shown that FW Tau C is a $\sim$0.1 $M_\sun$ star \citep{WS17}.

\footnotetext[1]{The GQ Lup data have been published in \cite{W17}, and we include them here for completeness.}

\begin{deluxetable*}{@{}lccccccccrc@{}}
\tablewidth{\linewidth}
\tablecaption{ALMA 1.3 mm Continuum Observations \label{tb0}}
\tablehead{
\colhead{\hspace{-32pt}Source} &
\colhead{Date} &
\colhead{$N_{\rm ant}$} &
\colhead{$L_{\rm baseline}$} &
\colhead{$T_{\rm obs}$} &
\colhead{Gain Cal.} &
\colhead{Flux Cal.} & 
\colhead{Bandpass Cal.} & 
\colhead{Beam} & 
\colhead{PA} & 
\colhead{rms} 
	\\
	\colhead{ }  &	
	\colhead{ }  &
	\colhead{ } &
	\colhead{ (m)} &	
	\colhead{(min)} &
	\colhead{} &
	\colhead{} &	
	\colhead{} &
	\colhead{($\arcsec$)} &
	\colhead{($\degr$)} &
	\colhead{($\mu$Jy beam$^{-1}$)} 
}
\startdata \vspace{2pt}
GQ Lup 		& 2015 Nov 01	& 41	& 85--14969 & $\sim$11 &	 J1534$-$3526	& J1337$-$1257  &	J1427$-$4206	&  $0.054\times0.031$ & 68.7	   & $\sim$40	\\
\vspace{2pt} 	 
DH Tau		& 2016 Sep 14	& 36	& 15--3247   & $\sim$13 & J0433$+$2905	& J0510$+$1800 &	J0510$+$1800	&  $0.286\times0.153$ & $-$18.8 & $\sim$43	\\
\vspace{2pt}
GSC 6214-210	& 2016 Sep 16	& 36	& 15--3143   & $\sim$12 & J1634$-$2058	& J1517$-$2422  &	J1517$-$2422	&  $0.202\times0.172$ & $-$2.3   & $\sim$30	\\
\vspace{2pt}
1RXS 1609	& 2016 Sep 16	& 36	& 15--3143   & $\sim$12 & J1634$-$2058	& J1517$-$2422  &	J1517$-$2422	&  $0.202\times0.176$ & $-$1.5   & $\sim$30	\\
CT Cha		& 2016 Sep 27	& 36 & 15--3247   & $\sim$14 & J1058$-$8003	& J1107$-$4449  &	J1107$-$4449	&  $0.259\times0.134$ & 16.5	   & $\sim$50
\enddata
\end{deluxetable*}

\begin{deluxetable*}{@{}lccccc@{}}
\tablewidth{\linewidth}
\tablecaption{Planet-mass Companions in Wide Orbits \label{tb1}}
\tablehead{
\colhead{} &
\colhead{GQ Lup B} &
\colhead{DH Tau B} &
\colhead{GSC 6214-210 B} &
\colhead{1RXS 1609 B} &
\colhead{CT Cha B} 
}
\startdata \vspace{2pt}
Mass ($M_{\rm Jup}$) 	& $\sim$10--40		& $\sim$15			& $\sim$15		& $\sim$13			& $\sim$20 		\\ \vspace{2pt}
SpT					& L1 $\pm$ 1		& M9.25 $\pm$ 0.25		& M9.5 $\pm$1 	& L2 $\pm$ 1 			& $\sim$M8		\\ \vspace{2pt}
log($L/L_\sun$)			& $-2.47\pm0.28$ 	& $-2.71\pm0.12$		& $-3.1\pm0.1$		& $-3.36\pm0.09$		& $-2.68\pm0.21$	\\ \vspace{2pt}
$T_{\rm eff}$ (K) 		& $2400\pm100$	& $2400\pm100$		& $2200\pm100$	& $2000\pm100$		& $2600\pm100$	\\ \vspace{2pt}
Age (Myr) 				& 2--5			& $\sim$2				& $\sim$10		& $\sim$10			& $\sim$2 		\\ \vspace{2pt}
$D$ (pc)				& $\sim$150		& $\sim$140			& $\sim$150		& $\sim$150			& $\sim$180		\\ \vspace{2pt}
$\rho$ (\arcsec) 		& 0.72			& 2.35 				& 2.17 			& 2.21				& 2.68			\vspace{2pt} \\
\hline \vspace{-2pt}
\\ \vspace{2pt}
Accretion and 		& H$\alpha$, Pa-$\beta$	& H$\alpha$, Pa-$\beta$	& H$\alpha$, Pa-$\beta$, Br-$\gamma$	& high $A_V$		& Pa-$\beta$\\ \vspace{2pt}
disk markers\tablenotemark{$\dagger$}		& red $K'-L'$		& red $K'-L'$		& red $K'-L'$			& red $K'-L'$ 		& high $A_V$\\ \vspace{2pt}
					& 				&& 24~\micron~excess 				&	24~\micron~excess				& \vspace{2pt} \\
\hline \vspace{-2pt}
\\ \vspace{2pt}
$\dot{M}$ ($M_{\rm Jup}$~yr$^{-1}$)  	&	$5.3\times10^{-7}$	&	$4.2\times10^{-9}$	&	$1.3\times10^{-8}$	&$\cdots$&$\cdots$
\\ \vspace{2pt}
3$\sigma$ Flux Limit ($\mu$Jy) 		&150 (880~\micron)	& 130 (1.3 mm) & 220 (880~\micron) & 90 (1.3 mm)& 150 (1.3 mm)\\ \vspace{2pt}
 		&120 (1.3 mm)	&& 90 (1.3 mm)	   & & \\ \vspace{2pt}
3$\sigma$ Dust Limit\tablenotemark{$\ddagger$} ($M_\oplus$)	& $<$0.04 & $<$0.07 & $<$0.05  & $<$0.06  & $<$0.14\\
References   			&1, 2, 3, 4		& 3, 4, 5, 6 		& 3, 4, 7, 8, 9		& 4, 7, 10		& 	11, 12
\enddata
\tablenotetext{$\dagger$}{The {\it Spitzer} 24~\micron~imaging is unresolved.}
\tablenotetext{$\ddagger$}{Since different authors have different assumptions of disk temperature, to facilitate comparisons we re-calculate dust mass limits using the reported 3$\sigma$ flux limits, assuming a disk temperature of 20 K for all objects, adopting dust opacities of 3.4 cm$^2$ g$^{-1}$ at 880~\micron~and 2.3 cm$^2$ g$^{-1}$ at 1.3 mm from \cite{B90}, and applying Equation \ref{eq1}.}
\tablerefs{(1) \cite{W17} and references therein, (2) \cite{M17}, (3) \cite{Z14}, (4) \cite{K14}, (5) \cite{Bonnefoy14} and references therein, (6) \cite{Wolff17}, (7) \cite{B13} and references therein, (8) \cite{B14}, (9) \cite{B15}, (10) \cite{W15b} and references therein, (11) \cite{W15a} and references therein, (12) \cite{Manara17}.}
\end{deluxetable*}

\begin{figure*}[t]
    \centering
    \includegraphics[width=\linewidth]{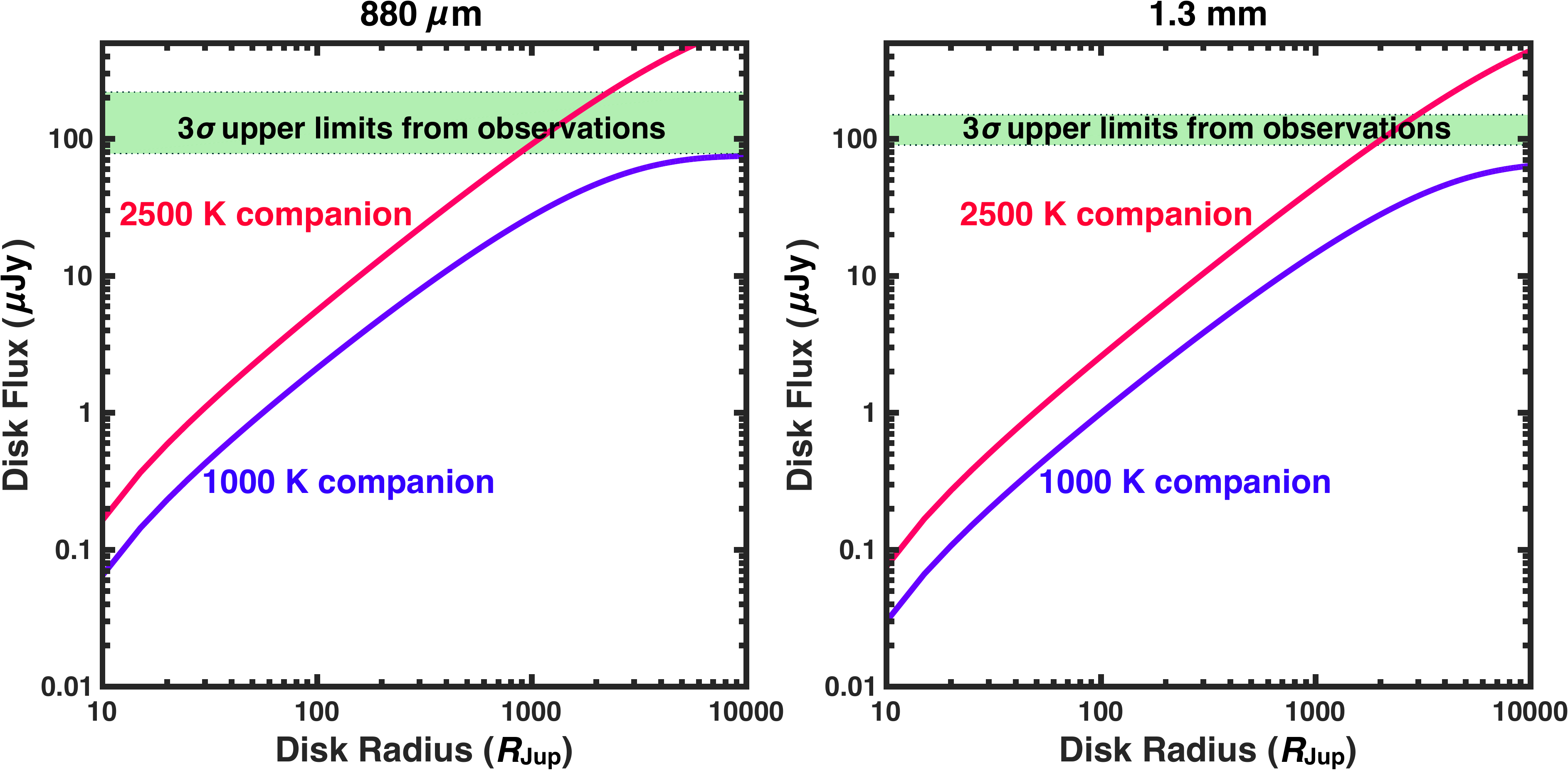}
    \caption{Disk flux as a function of disk radius for 1000 K and 2500 K companions under the optically thick approximation. ALMA 880 \micron~and 1.3 mm flux limits suggest that compact optically-thick disks are probably smaller than 1000 $R_{\rm Jup}$, or $\sim$0.5 au.}
    \label{fig1}
\end{figure*}

\section*{\textbf {\normalsize2. O\lowercase{ptically-thin} D\lowercase{ust} \lowercase{and} T\lowercase{imescale} P\lowercase{roblem}}}
Assuming optically thin dust, one can estimate the dust mass from \cite{H83}, 
\begin{equation}\label{eq1}
M_{\rm dust} = \frac{F_{\nu}~D^2}{\kappa_{\nu}~B_{\nu}(T)},
\end{equation}
where $F_{\nu}$ is the observed flux, $D$ is the distance to the source, $\kappa_{\nu}$ is the dust opacity, and $B_{\nu}$ is the Planck function. In Table \ref{tb1}, we list the physical properties, disk evidence, 3$\sigma$ radio flux limits (from our observations and the literature), and the corresponding dust mass limits for 5 planetary-mass companions in our survey. We notice that the $\sim$5 $M_{\rm Jup}$ planet 2M1207 b has a 880~\micron~flux limit of $\sim$80 $\mu$Jy recently measured by \cite{R17}. Hence, current ALMA observations have reached 3$\sigma$ sensitivities of 80 to 220 $\mu$Jy at 880~\micron~and 90 to 150 $\mu$Jy at 1.3 mm for wide companions. These flux upper limits translate to a dust mass of 0.002 to 0.14 $M_\oplus$ (0.2 to 11.4 $M_{\rm moon}$) assuming a characteristic disk temperature of 20 K. Since many wide companions have multiple features suggestive of accretion disks, it is surprising and puzzling that radio observations have placed such a strong constraint on the amount of dust. The small amount of dust in turn implies a very short disk lifetime due to accretion. For instance, GQ Lup B has an accretion rate of $5\times10^{-7}~M_{\rm Jup}$ yr$^{-1}$~\citep{Z14}, yet an optically thin disk has $<$0.04 $M_\oplus$ of dust \citep{M17}. Hence, GQ Lup B's accretion disk could potentially disappear in $\sim$20 kyr, much shorter than the 2--5 Myr age of the system. If the actual gas-to-dust ratio is lower than the canonical value of 100, as hinted by recent ALMA surveys on T Tauri disks (e.g., \citealt{A16,E16}), then GQ Lup B's disk would be depleted even faster.

This dramatic timescale difference seems to suggest that we are observing planet-mass companions at a very special time close to the very end of accretion, which is {\it a priori} unlikely. We note that at an age of $\sim$10 Myr, GSC 6214-210 B still has lines of evidence arguing for active accretion (see Table \ref{tb1}). This implies that disks around planet-mass companions can indeed survive for a long period, presumably longer than the average lifetime of protoplanetary disks ($\sim$5 Myr; \citealt{H01}).

\section*{\textbf {\normalsize3. C\lowercase{ompact} O\lowercase{ptically-thick} D\lowercase{isk}}}
Low radio flux does not necessarily mean low dust mass. A more natural explanation is that their disks may have the mass needed to sustain accretion and satellite formation for a few to even $>$10 Myr ($>$0.1 $M_{\rm Jup}$ or even $>$1 $M_{\rm Jup}$), but they appear faint in (sub)millimeter because the disks are compact. This is also hinted by infrared observations as some companions have near- or mid-infrared excesses likely from hot inner disks (Table \ref{tb1}), implying that the lack of (sub)millimeter detections requires small disk radii.

Compact dust continuum emission is optically thick, rather than optically thin as usually assumed. We note that similar thoughts have been discussed and applied to derive an upper limit on disk size \citep{B15,M17,R17,Wolff17}. Here, we further study the brightness and implications of compact optically-thick disks.

Assuming disk heating is dominated by the irradiation from the companion, we can roughly estimate the brightness of an optically-thick disk following \cite{P81}. The temperature profile is
\begin{equation}\label{eq2}
T(r) = T_\star~(r/R_\star)^{-3/4}~\big(1-\sqrt{R_\star/r}\big)^{1/4},
\end{equation}
where $T_\star$ and $R_\star$ are the effective temperature and radius of the central object, respectively. For $r \gg R_\star$, $T(r)\propto r^{-3/4}$, which is similar to the profile seen in some simulated circumplanetary disks (e.g., \citealt{AB09}). We can calculate the disk flux $F_\nu$ by integrating the Planck function $B_\nu$ over solid angle:
\begin{equation}\label{eq3}
F_\nu =  \int B_\nu(T(r))\,d\Omega = \frac{2\pi}{D^2} \int^{R_{\rm max}}_{R_{\rm min}} B_\nu(T(r))\,r\,dr,
\end{equation}
where $R_{\rm min}$ and $R_{\rm max}$ are the disk inner and outer radii, respectively. At long wavelengths, $B_\nu~\propto~T$, so $F_\nu$ approximately scales as $R_{\rm max}$.

We adopt $R_\star=2.5~R_{\rm Jup}$, which is typical for young substellar objects (e.g., \citealt{Baraffe15}), $R_{\rm min} = 2~R_\star$, $R_{\rm max}$ = 10 to $10^4~R_{\rm Jup}$ ($\sim$0.005 to 5 au), and $D$ = 150 pc as most nearby star-forming regions are approximately at that distance. Finally, we choose $T_\star$ = 2500 K and 1000 K as most companions have spectral types from late M to mid L (Table \ref{tb1}).

Figure \ref{fig1} shows the disk radius versus the 880~\micron~(ALMA band 7) and 1.3 mm (ALMA band 6) disk fluxes given by Equation \ref{eq3}. We also label the 3$\sigma$ flux upper limits from ALMA observations. We can see that compact optically-thick disks are indeed very faint at radio wavelengths --- current observational constraints suggest that they are fainter than $\sim$100 $\mu$Jy. As a result, they cannot be as large as 1/3 of the Hill radius ($\sim$5 to 30 au in radius), or we would have easily detected them with ALMA. Instead, Figure \ref{fig1} suggests that they are probably smaller than 1000 $R_{\rm Jup}$, or $\sim$0.5 au. Such small optically-thick disks can still contain all the gas and dust needed for a few Myr of steady accretion onto the companion, hence solving the lifetime problem created by the ALMA non-detections.

\begin{figure*}[t]
    \centering
    \includegraphics[width=\linewidth]{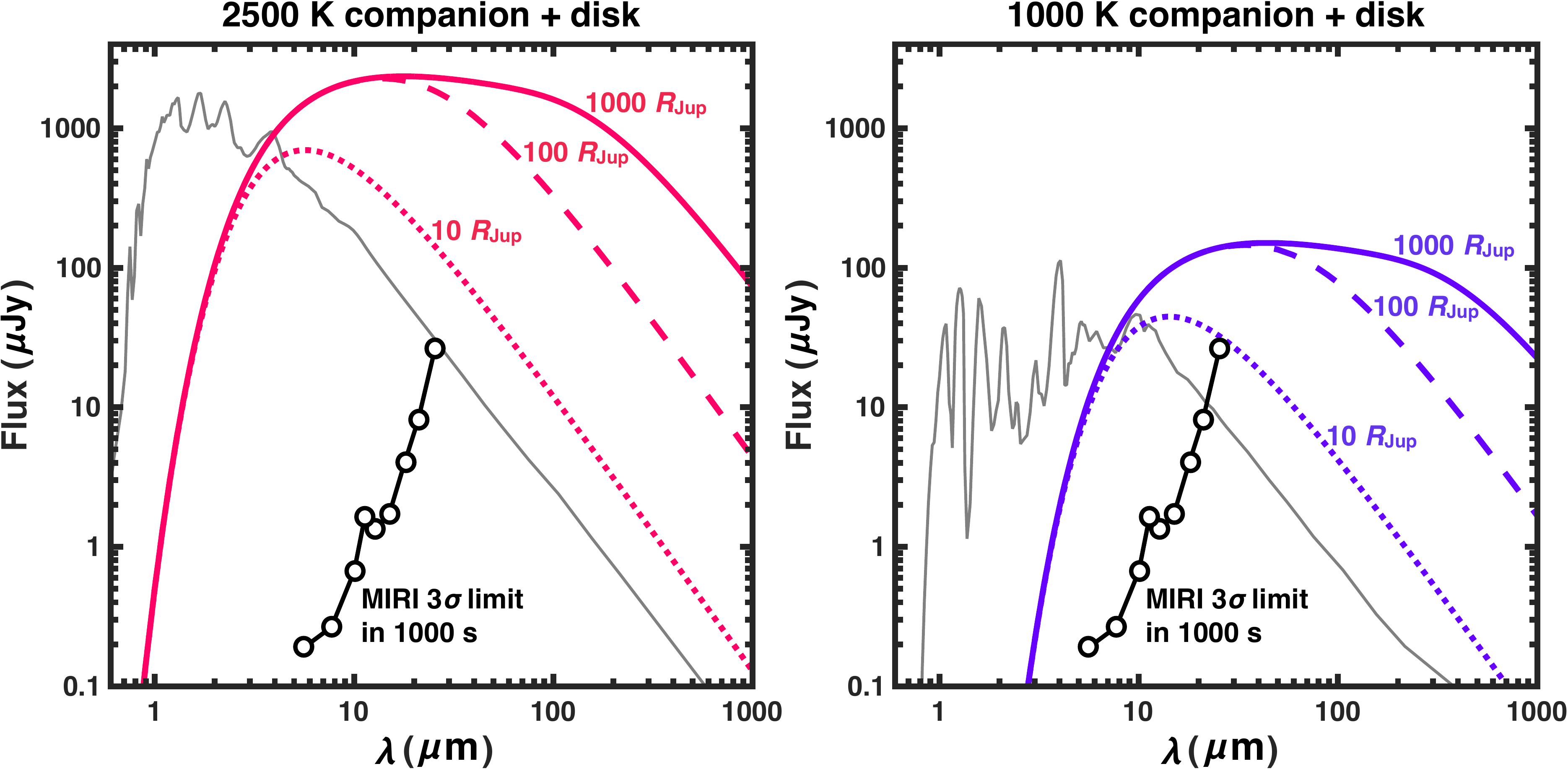}
    \caption{Spectral energy distributions for compact optically-thick disks with radii of 10, 100, and 1000 $R_{\rm Jup}$. The 2500 K and 1000 K BT-Settl models \citep{A11} are shown in gray. The 1000 s MIRI photometric detection limits suggest that {\it JWST} has the potential to constrain disk sizes for most wide-orbit planet-mass companions.
}
    \label{fig2}
\end{figure*}

\section*{\textbf {\normalsize4. I\lowercase{mplications}}}
If these planet-mass companions do indeed have compact optically-thick disks, then our calculations have a few implications:

1. ALMA may not be the ideal instrument to detect and characterize circumplanetary disks. As Figure \ref{fig1} shows, disk flux roughly scales with disk size, so detecting a disk that is 10 times smaller in size would require $\sim$100 times longer integration to reach the same signal-to-noise ratio. Therefore, an unrealistically long integration is required for ALMA to reach an rms of 1 $\mu$Jy in order to detect disks smaller than 100 $R_{\rm Jup}$.

2. Mid-infrared observations, which probe the peak of the spectral energy distribution (SED) of compact disks, may be more favorable for constraining disk sizes in order to compare with theories. The average temperature of compact optically-thick disks can be higher than that of T Tauri disks ($\sim$25 K; e.g., \citealt{A13}). The area-weighted disk temperatures, $T_{\rm disk} = \int^{R_{\rm max}}_{R_{\rm min}} T(r)\,2\pi r\,dr / (\pi R_{\rm max}^2-\pi R_{\rm min}^2)$, for disk radii of 10--1000 $R_{\rm Jup}$ ($\sim$0.005--0.5 au) range from $\sim$40 K to $\sim$880 K for a 2500 K companion, and $\sim$20 K to $\sim$350 K for a 1000 K companion. As a result, compact disks are bright in the mid- to far-infrared. Figure \ref{fig2} shows the SEDs derived from Equation \ref{eq3} for 10, 100, and 1000 $R_{\rm Jup}$ compact disks. It is evident that the peak of SED is size-dependent, ranging from $\sim$10 to $\sim$100~\micron. We also plot the 3$\sigma$ point source detection limits for 1000 s integration time for the Mid-Infrared Instrument (MIRI) on the {\it James Webb Space Telescope (JWST)}. MIRI's superior sensitivity at 10 to 25~\micron~will be able to constrain disk sizes for most young planetary-mass companions.

3. The origin of planetary-mass companions needs further exploration. Non-detections of wide companion disks are in stark contrast to recent studies which reveal that brown dwarf disks can have 0.1--1 $M_\oplus$ of dust and tens of au in radius (e.g., \citealt{Ricci14,V16,H17}). Moreover, the disk around the 12 $M_{\rm Jup}$ free-floating planet OTS 44 has recently been imaged with ALMA and is found to have 0.1--0.6 $M_\oplus$ of dust \citep{B17}. Hence, even though planet-mass companions share similar masses to field brown dwarfs and free-floating planets, they probably have a different formation pathway that involves dynamical encounters with other massive bodies to truncate their disks and also scatter them to outer orbits (e.g., \citealt{B03}). However, searches for inner massive bodies that can serve as the scatterers seem to disagree with the scattering scenario but favor the in situ formation via disk fragmentation or prestellar core collapse \citep{B16}. 

4. Moons could be hard to form in compact optically-thick disks in the first few Myr because of the high disk temperature. In the inner disk, the temperature can be even higher than the dust sublimation temperature. As a result, satellite formation might have to wait until the companion cools off, or only happen in the outer disk where the temperature is lower (e.g., \citealt{Z16,S17}). 

\acknowledgements
We thank the referee for helpful comments. Y.-L.W. is supported by the NSF AAG award and the TRIF fellowship. This paper makes use of the following ALMA data: ADS/JAO.ALMA\#2015.1.00773.S. ALMA is a partnership of ESO (representing its member states), NSF (USA) and NINS (Japan), together with NRC (Canada), NSC and ASIAA (Taiwan), and KASI (Republic of Korea), in cooperation with the Republic of Chile. The Joint ALMA Observatory is operated by ESO, AUI/NRAO, and NAOJ. The National Radio Astronomy Observatory is a facility of the National Science Foundation operated under cooperative agreement by Associated Universities, Inc.


\begin{thebibliography}{}
\bibitem[Allard et al.(2011)]{A11}Allard, F., Homeier, D., \& Freytag, B. 2011, in ASP Conf. Ser. 448, XVI Cambridge Workshop on Cool Stars, Stellar Systems, and the Sun, ed. C. Johns-Krull, M. K. Browning, \& A. A. West (San Francisco, CA: ASP), 91
\bibitem[Allers \& Liu(2013)]{AL13}Allers, K. N., \& Liu, M. C. 2013, \apj, 772, 79
\bibitem[Andrews et al.(2013)]{A13}Andrews, S. M., Rosenfeld, K. A., Kraus, A. L., \& Wilner, D. J. 2013, \apj, 771, 129
\bibitem[Ansdell et al.(2016)]{A16}Ansdell, M., Williams, J. P., van der Marel, N., et al. 2016, \apj, 828, 46
\bibitem[Ayliffe \& Bate(2009)]{AB09}Ayliffe, B. A., \& Bate, M. R. 2009, \mnras, 397, 657
\bibitem[Bailey et al.(2013)]{B13}Bailey, V., Hinz, P. M., Currie, T., et al. 2013, \apj, 767, 31
\bibitem[Baraffe et al.(2015)]{Baraffe15}Baraffe, I., Homeier, D., Allard, F., \& Chabrier, G. 2015, \aap, 577, A42
\bibitem[Barman et al.(2011)]{Barman11}Barman, T. S., Macintosh, B., Konopacky, Q. M., \& Marois, C. 2011, \apjl, 735, L39
\bibitem[Bate et al.(2003)]{B03}Bate, M. R., Bonnell, I. A., \& Bromm, V. 2003, \mnras, 339, 577
\bibitem[Bayo et al.(2017)]{B17}Bayo, A., Joergens, V., Liu, Y., et al. 2017, \apjl, 841, L11
\bibitem[Beckwith et al.(1990)]{B90}Beckwith, S. V. W., Sargent, A. I., Chini, R. S., \& Guesten, R. 1990, \aj, 99, 924
\bibitem[Biller et al.(2014)]{Biller14}Biller, B. A., Males, J. R., Rodigas, T. J., et al. 2014, \apjl, 792, L22
\bibitem[Bonnefoy et al.(2014)]{Bonnefoy14}Bonnefoy, M., Chauvin, G., Lagrange, A.-M., et al. 2014, \aap, 562, A127
\bibitem[Bowler et al.(2015)]{B15}Bowler, B. P., Andrews, S. M., Kraus, A. L., et al. 2015, \apjl, 805, L17
\bibitem[Bowler et al.(2014)]{B14}Bowler, B. P., Liu, M. C., Kraus, A. L., \& Mann, A. W. 2014, \apj, 784, 65
\bibitem[Bowler et al.(2011)]{Bowler11}Bowler, B. P., Liu, M. C., Kraus, A. L., Mann, A. W., \& Ireland, M. J. 2011, \apj, 743, 148
\bibitem[Bryan et al.(2016)]{B16}Bryan, M. L., Bowler, B. P., Knutson, H. A., et al. 2016, \apj, 827, 100
\bibitem[Brandt et al.(2014)]{Brandt14}Brandt, T. D., McElwain, M. W., Turner, E. L., et al. 2014, \apj, 794, 159
\bibitem[Caceres et al.(2015)]{C15}Caceres, C., Hardy, A., Schreiber, M. R., et al. 2015, \apjl, 806, L22
\bibitem[Eisner(2015)]{E15}Eisner, J. A. 2015, \apjl, 803, L4
\bibitem[Eisner et al.(2016)]{E16}Eisner, J. A., Bally, J. M., Ginsburg, A., \& Sheehan, P. D. 2016, \apj, 826, 16
\bibitem[Follette et al.(2017)]{F17}Follette, K. B., Rameau, J., Dong, R., et al. 2017, \aj, 153, 264
\bibitem[Haisch et al.(2001)]{H01}Haisch, K. E., Lada, E. A., \& Lada, C. J. 2001, \apj, 553, L153
\bibitem[Hendler et al.(2017)]{H17}Hendler, N., Mulders, G. D., Pascucci, I., et al. 2017, \apj, 841, 116
\bibitem[Hildebrand(1983)]{H83}Hildebrand, R. H. 1983, QJRAS, 24, 267
\bibitem[Isella et al.(2014)]{I14}Isella, A., Chandler, C. J., Carpenter, J. M., P\'{e}rez, L. M., \& Ricci, L. 2014, \apj, 788, 129
\bibitem[Kraus et al.(2015)]{K15}Kraus, A. L., Andrews, S. M., Bowler, B. P., et al. 2015, \apjl, 798, L23
\bibitem[Kraus et al.(2014)]{K14}Kraus, A. L., Ireland, M. J., Cieza, L. A., et al. 2014, \apj, 781, 20
\bibitem[MacGregor et al.(2017)]{M17}MacGregor, M. A., Wilner, D. J., Czekala, I., et al. 2017, \apj, 835, 17
\bibitem[Males et al.(2014)]{Males14}Males, J. R., Close, L. M., Morzinski, K., et al. 2014, \apj, 786, 32
\bibitem[Manara et al.(2017)]{Manara17}Manara, C. F., Prusti, T., Voirin, J., \& Zari, E. 2017, arXiv:1707.03179
\bibitem[Nielsen \& Close(2010)]{NC10}Nielsen, E. L., \& Close, L. M. 2010, \apj, 717, 878
\bibitem[Pringle(1981)]{P81}Pringle, J. E. 1981, \araa, 19, 137
\bibitem[Quillen \& Trilling(1998)]{QT98}Quillen, A. C., \& Trilling, D. E. 1998, \apj, 508, 707
\bibitem[Ricci et al.(2017)]{R17}Ricci, L., Cazzoletti, P., Czekala, I., et al. 2017, \aj, 154, 24
\bibitem[Ricci et al.(2014)]{Ricci14}Ricci, L., Testi, L., Natta, A., et al. 2014, \apj, 791, 20
\bibitem[Sallum et al.(2015a)]{S15a}Sallum, S., Eisner, J. A., Close, L. M., et al. 2015, \apj, 801, 85
\bibitem[Sallum et al.(2015b)]{S15b}Sallum, S., Follette, K. B., Eisner, J. A., et al. 2015, \nat, 527, 342
\bibitem[Shabram \& Boley(2013)]{SB13}Shabram, M., \& Boley, A. C. 2013, \apj, 767, 63
\bibitem[Szul\'{a}gyi(2017)]{S17}Szul\'{a}gyi, L. 2017, \apj, 842, 103
\bibitem[van der Plas et al.(2016)]{V16}van der Plas, G., M\'{e}nard, F., Ward-Duong, K., et al. 2016, \apj, 819, 102
\bibitem[Wolff et al.(2017)]{Wolff17}Wolff, S. G., M\'{e}nard, F., Caceres. C., et al. 2017, \aj, 154, 26
\bibitem[Wu et al.(2015a)]{W15a}Wu, Y.-L., Close, L. M., Males, J. R., et al. 2015a, \apj, 801, 4
\bibitem[Wu et al.(2015b)]{W15b}Wu, Y.-L., Close, L. M., Males, J. R., et al. 2015b, \apjl, 807, L13
\bibitem[Wu et al.(2017)]{W17}Wu, Y.-L., Sheehan, P. D., Males, J. R., et al. 2017, \apj, 836, 223
\bibitem[Wu \& Sheehan(2017)]{WS17}Wu, Y.-L., \& Sheehan, P. D. 2017, \apjl, 846, L26
\bibitem[Zhou et al.(2014)]{Z14}Zhou, Y., Herczeg, G. J., Kraus, A. L., Metchev, S., \& Cruz, K. L. 2014, \apjl, 783, L17
\bibitem[Zhu(2015)]{Z15}Zhu, Z. 2015, \apj, 799, 16
\bibitem[Zhu et al.(2016)]{Z16}Zhu, Z., Ju, W., \& Stone, J. M. 2016, \apj, 832, 193
\end{thebibliography}
\end{document}